\documentclass[superscriptaddress,aps,prl,twocolumn,showpacs,preprintnumbers,amsmath,amssymb,floatfix,groupedaddress]{revtex4}

\usepackage{graphicx}
\usepackage{dcolumn}
\usepackage{bm}
\usepackage{color}

\begin{document}

\title{Why the Water Bridge does not collapse}

\author{Artem A. Aerov}
\affiliation{Physics Department, Moscow State University, Moscow
119991, Russia}

\date{\today}

\begin{abstract}

In 2007 an interesting phenomenon was discovered: a thread of
water, the so-called water bridge (WB), can hang between two glass
beakers filled with deionized water if voltage is applied to them.
We analyze the available explanations of the WB stability and
propose a completely different one: the force that supports the WB
is the surface tension of water and the role of electric field is
not to allow the WB to reduce its surface energy by means of
breaking into separate drops.
\end{abstract}

\pacs{47.55.nk, 47.20.Dr, 77.22.-d, 47.65.-d}

\maketitle

\begin{figure}[t]
\begin{center}
\includegraphics[width = \columnwidth]{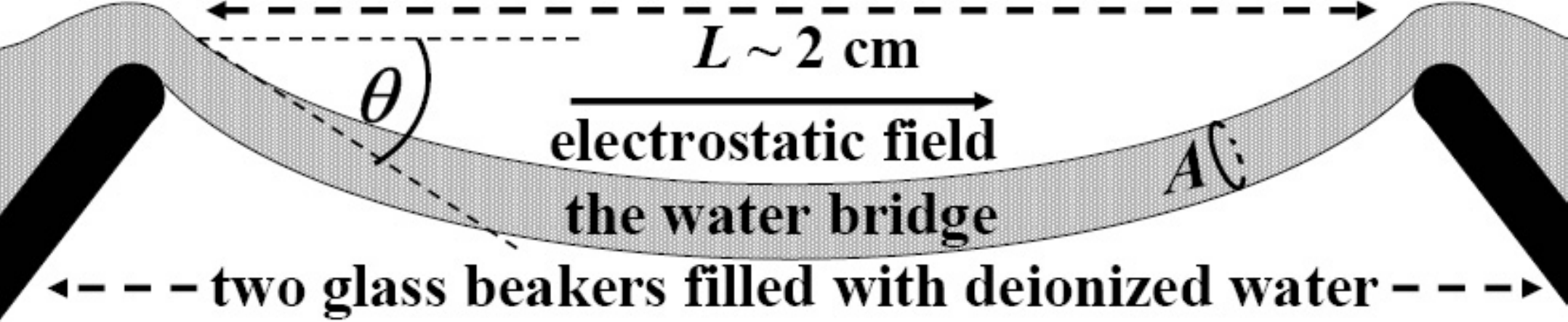}
\caption{Schematic illustration of the water bridge. \label{fig1}}
\end{center}
\end{figure}

After the WB (see Fig.~\ref{fig1}) was rediscovered in 2007
\cite{Fuchs_first} (it had been first time observed in
1893~\cite{First}) it immediately captured attention and even
entered some TV shows because the experiment is easy to reproduce
and it can be treated as an evidence of some unique properties of
water. What keeps WB stable against gravity? The first thing one
can suppose is that the water in WB has properties similar to
those of a polymer melt; i.e. in the electric field water
molecules are arranged in quasi polymer chains that play the role
of the WB load-carrying structure~\cite{Nishimi2009}. It has been
also supposed that hydrogen bonds are the driving force of WB
formation~\cite{Cramer2008}. But in the computer simulation
carried out in the work~\cite{Cramer2008} the WB consisted of only
$10^3$ molecules and it could be formed if the electric field was
at least $\approx 10^3$ times stronger than the one necessary for
formation of macroscopic WBs in real
experiments~\cite{Fuchs_first,Fuchs_neutron,Dynamics,Saija_2010_model,Ponterio_2010_raman,Marin_2010}.
Some attempts have been made to reveal a specific structure of WB
by means of neutron scattering and Raman
scattering~\cite{Fuchs_neutron,Ponterio_2010_raman}, but no
exhaustive explanation of the WB stability has been found on this
way. An interesting feature of WB is the complicated spiral flow
of water and formation of tiny bubbles inside it~\cite{Dynamics}.
But it has not been proved yet that the dynamics of WB can be
related to its stability. It has been even supposed that WB
stability against gravity is a quantum effect~\cite{Quantum}.

\begin{figure}[t]
\begin{center}
\includegraphics[width = \columnwidth]{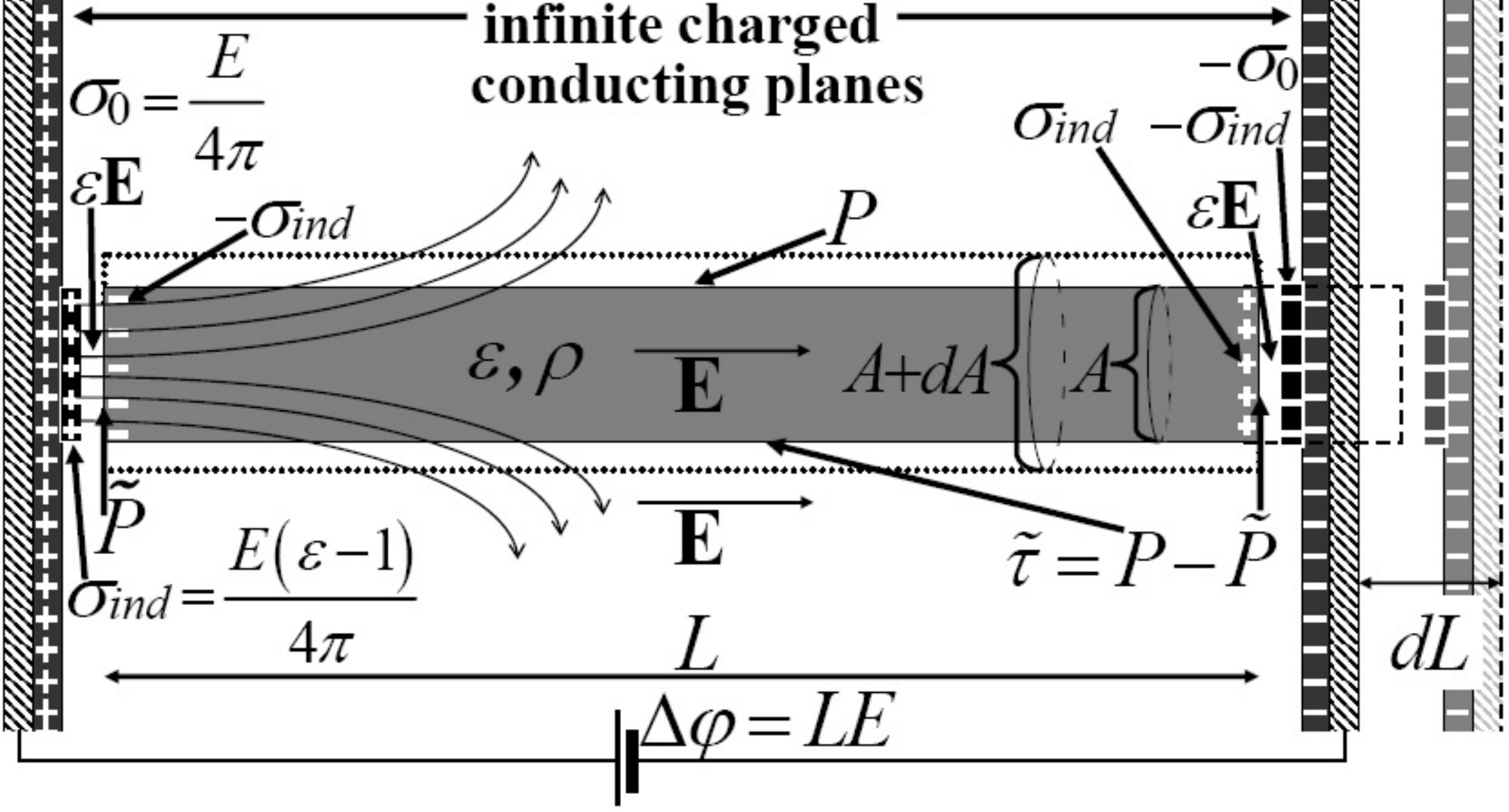}
\caption{Schematic illustration of a DL cylinder in a uniform
electrostatic field between two infinite conducting planes.
\label{fig2}}
\end{center}
\end{figure}

However, the best explanation of a newly discovered phenomenon is
the simplest one based on well known formulas. It has been stated
already~\cite{Widom_2009, Saija_2010_model} that not specific
properties of water but just its high dielectric permittivity is
likely to be the reason of the WB phenomenon. The convincing
evidence of the statement is the "water bridge" ({\it dielectric
liquid bridge}~(DB)) formation of another low molecular polar
dielectric liquid (DL): glycerine~\cite{Marin_2010}. A good hint
for the discovery with glycerine is the necessity to deionize
water for forming WB. How can the high dielectric permittivity of
a DL cause the DB stability? It is straightforward to assume that
DB is kept stable against gravity by tension as a hanging flexible
cable~\cite{Widom_2009}, the tension being somehow produced by
electric field. Let us imagine a DL cylinder in a uniform
electrostatic field (EF)~${\bf E}$ parallel to its axis. This is
possible if the cylinder bases touch two infinite conducting
planes to which voltage $\Delta \varphi$ is applied~(see
Fig.~\ref{fig2}). To simplify the explanation we have depicted in
Fig.~\ref{fig2} gaps between the bases and the planes. The gaps
are supposed to be infinitely thin, the pressure produced by the
bases on the gaps is actually the pressure produced by the bases
on the planes. The cylinder is the simplest model of DB.
EF exerts pressure on a perpendicular to it DL
interface~\cite{Landafshiz,BeckerSauter}. This pressure~$\tilde P$
on the cylinder bases is claimed in Ref.~\cite{Saija_2010_model}
to be the reason of the tension, which does not allow gravity to
rupture DB. But this explanation of the stability is not correct,
at least because EF parallel to a DL surface also produces a
pressure~$P$ on it~\cite{Landafshiz,BeckerSauter} (see
Fig.~\ref{fig2}) and only the effective DB tension~$\tilde
\tau$~\cite{Widom_2009} could be the reason of the stability.
Assuming the pressure~$P$ to be the tension holding DB, which is
done in Ref.~\cite{Marin_2010}, is also a strange idea. More
specifically, if~$L$ and~$A$ are the length and the cross-section
area of the cylinder, the work consumed for its small elongation
is $dW = \tilde \tau A dL \label{eq1}$. At the same time $dW =
-\tilde P A dL - P L dA \label{eq2}$. In the first approximation
the DL is incompressible, i.e.~$AdL+LdA=0$. Therefore, $\tilde
\tau = P -\tilde P$. A positive value of~$\tilde \tau$ could
explain the DB stability. This explanation is proposed in
Ref.~\cite{Widom_2009} and seems at first to be the most
consistent one. But let us calculate~$\tilde \tau$ carefully.

First, something strange is written in the Introduction of
Ref.~\cite{Widom_2009}. Namely, the normal component of the
electrostatic displacement~${\bf D}=\varepsilon{\bf E}$
($\varepsilon$ is the dielectric constant of the medium) is
actually not discontinuous on a dielectric/dielectric interface.
Besides, positive rather than negative charge is induced on the
dielectric cylinders base of the "electric field arrow"~(the right
base of the cylinder in Fig.~\ref{fig2}). Moreover, it is
essentially senseless to calculate the cylinder tension produced
by the action of the external \emph{uniform} electric field~${\bf
E}$ on this charge, because the field acts with a zero force on
each dipole molecule of a dielectric; charges induced on a
dielectric/dielectric interface are not free charges. $P$
and~$\tilde P$ are obtained in Ref.~\cite{Widom_2009} by using the
expression for the {\it Maxwell stress tensor} (see Eqs.~(15.9),
(36.6), or~(14.9) in
Refs.~\cite{Landafshiz},~\cite{BeckerSauter},~\cite{Denisov}
respectively) in the interior of the DL cylinder. But to obtain
the electrostatic pressure on the interface of two DLs one must
subtract the Maxwell stresses on both sides of the
interface~\cite{BeckerSauter,Denisov}. As concerns DB, the Maxwell
tension of vacuum is to be subtracted in the beginning from~$P$
and~$\tilde P$ and not in the end from~$\tilde \tau$ as has been
done in Ref.~\cite{Widom_2009}. Then the general expression for
the pressures is the Eq.~(36.11) in Ref.~\cite{BeckerSauter}, and
the pressure on air/DL interface is expressed by Eq.~(15.11) in
Ref.~\cite{Landafshiz}:
\begin{eqnarray}
P = \frac{\varepsilon -1}{8\pi}E^2- \frac{\rho
E^2}{8\pi}\left(\frac{\partial \varepsilon }{\partial
\rho}\right)_T =\tilde P-\frac{{(\varepsilon
-1})^2}{8\pi}E^2\label{eq4}
\end{eqnarray}
where $\rho$ and $T$ are respectively the DL density and
temperature, and the normal is directed outside DL.

Let us derive Eq.~(\ref{eq4}). It follows from the EF boundary
conditions that 1) EF is same inside and outside the cylinder (see
Fig.~\ref{fig2}) 2) the surface densities of charges induced on
the cylinder bases are~$\mp\sigma_{ind}$ while the densities on
the corresponding adjacent areas of the conducting planes
are~$\pm\left(\sigma_{0}+\sigma_{ind}\right)$, and the densities
on the corresponding rest parts of the planes are~$\pm\sigma_{0}$.
If the cylinder cross-section area is isothermally increased at
constant voltage $\Delta \varphi$ by~$dA$ while the length $L$ is
kept constant, the EF energy $1/(8\pi)\int \varepsilon E^2
d^3\textbf{r}$ is changed by $dU^{\bot}$ and the voltage does the
work~$dW^{\bot}$:
\begin{eqnarray}
dU^{\bot} = \frac{dW^{\bot}}{2} =\frac{\varepsilon -1}{8\pi} E^2
LdA - \frac{ E^2}{8\pi}\rho\left(\frac{\partial \varepsilon
}{\partial \rho}\right)_T LdA
\end{eqnarray}
where it is taken into account that the cylinders dielectric
permittivity and volume are changed due to the stretching, and the
change of the permittivity and the cross-section area leads also
to the change of the planes charges. It follows from the energy
conservation law that $PLdA=dW^{\bot}-dU^{\bot}$, which gives the
first part of Eq.~(\ref{eq4}). Let us now suppose that the planes
are isothermally moved apart by~$dL$ at constant~$\Delta \varphi$,
and the cylinders length~$L$ is increased respectively by~$dL$
while the cross-section area~$A$ is kept constant. For the part of
the system outer to the cylinder the difference between the
voltage work and the EF energy increase gives the mutual planes
coulomb attraction force existing independently of the cylinder.
In the cylinder part the EF energy is changed by $dU^{\|}$ and the
voltage does the work~$dW^{\|}$:
\begin{eqnarray}
dU^{\|} = \frac{dW^{\|}}{2}=-\frac{\varepsilon}{8\pi} E^2 AdL -
\frac{ E^2}{8\pi}\rho\left(\frac{\partial \varepsilon }{\partial
\rho}\right)_T AdL \label{eq8}
\end{eqnarray}
where it is taken into account that the cylinders dielectric
permittivity and volume are changed, and also the EF is decreased
by~$EdL/L$. In this case one should be careful to avoid a mistake:
not only the pressure~$\tilde P$ does the work but also the
coulomb attraction of the charges on the planes (see also the end
of \S~37 in Ref.~\cite{BeckerSauter}):
\begin{eqnarray}
dW^{\|}-dU^{\|}=\left(\tilde
P-\left(\sigma_{0}+\sigma_{ind}\right) \varepsilon E /2 \right)AdL
\label{eq9}
\end{eqnarray}
where~$\varepsilon E$ is the EF in the gaps between the cylinder
bases and the planes. Eqs. (\ref{eq8}) and (\ref{eq9}) give the
second part of Eq.~(\ref{eq4}).

So, DB pushes apart glass beakers with the pressure~$-\tilde
\tau=\left(\varepsilon-1\right)^2 E^2/(8\pi)$ (see
Eq.~(\ref{eq4})) instead of pulling them towards each other as a
suspension bridge pulls its pillars. Hence, the explanation
proposed in Ref.~\cite{Widom_2009} is not adequate. But the not
existing tension of DB derived in Ref.~\cite{Widom_2009} has an
annoying feature: it is in some cases close in value to the one
that really would hold DB. For this reason the theory in
Ref.~\cite{Widom_2009} even finds experimental
"corroborations"~\cite{Marin_2010}. We propose another
experimental verification: if the theory is correct, a DB that is,
say, two times longer ($L\sim 3.5-4.5~cm$) is possible in a two
times stronger EF. Seems to be not the case. What tension holds DB
then? The EF~$\varepsilon E/2$, in the gap, say, at the left plane
(minus the plane EF)~(see Fig.~\ref{fig2}) can be presented as the
sum of the values~$E/2$ and~$\left(\varepsilon-1\right)E/2$.
The~$E/2$ is produced by the uniform charge density~$-\sigma_{0}$
of the right plane, i.e.~the attraction of the left plane by this
EF is the force produced by the right plane. If~$L\ll\sqrt{A}$,
EFs of the charges induced on the cylinder bases cancel out in the
gap, and the field~$\left(\varepsilon-1\right)E/2$ is produced
only by the charge density~$-\sigma_{ind}$ on the right plane. In
this case the DL "pancake" would really only exert pressure on the
planes. But in the case of a DB,~$L\gg\sqrt{A}$, the field of the
$-\sigma_{ind}$~circle on the right plane vanishes in the left
gap, i.e.~the EF~$\left(\varepsilon-1\right)E/2$ is produced by
the cylinder. The total coulomb interaction of the cylinder and
the charge~$\sigma_{0}$ on the left plane is equal to zero,
because the uniform charge~$\sigma_{0}$ produces a uniform EF and
the total charge of the cylinder is zero. As to the attraction
between the~$\sigma_{ind}$ circle of the left plane and the long
cylinder, it is equal to the attraction~$2\pi\sigma_{ind}^2A$
between the circle and the opposite to it charge induced on the
left cylinder base because the charge on the right base is far
off. The coulomb force subtracted in the right part of the
Eq.~(\ref{eq9}) consists of the attraction~$A\varepsilon
E^2/(8\pi)$ between the~$\sigma_{0}+\sigma_{ind}$~circle on the
left plane and the uniform charge $-\sigma_{0}$ on the right
plane, of the attraction~$A(\varepsilon-1) E^2/(8\pi)$ between the
uniform charge~$\sigma_{0}$ on the left plane and
the~$-\sigma_{ind}$ circle on the right plane, and of the
attraction~$2\pi\sigma_{ind}^2A=A(\varepsilon-1)^2 E^2/(8\pi)$
between the $\sigma_{ind}$~circle on the left plane with the
cylinder. DB not only exerts pressure~$-\tau_{eff}$ on a plane but
also pulls it by EF. To which part of DB is the latter force
applied? The force exerted by EF~${\bf E}$ on a small volume of
dielectric is~$({\bf P}\nabla){\bf E}$, where~${\bf P}$ is the
dipole moment of the volume. The dipole moment density of the
cylinder is uniform:~$p=(\varepsilon-1)E/(4\pi)$. The
EF~$E_{\sigma_{ind}}$ of the left plane~$\sigma_{ind}$ circle is
equal to~$(\varepsilon-1)E/2$ on the left cylinder base, and it is
equal to zero on the right base: all the EF lines go out through
the cylinder lateral surface~(see Fig.~\ref{fig2}). Hence, the
total force~$\int ({\bf p}\nabla){\bf E_{\sigma_{ind}}}d^3r$ is
equal to~$2\pi\sigma_{ind}^2A$ and it is applied to the left
segment of the cylinder where the lines cross its lateral surface.
The segment characteristic length is~$\sqrt{A}$. The DB
pressure~$-\tau_{eff}$ on a plane and its attraction of it by EF
cancel out.

\begin{figure}[t]
\begin{center}
\includegraphics[width = \columnwidth]{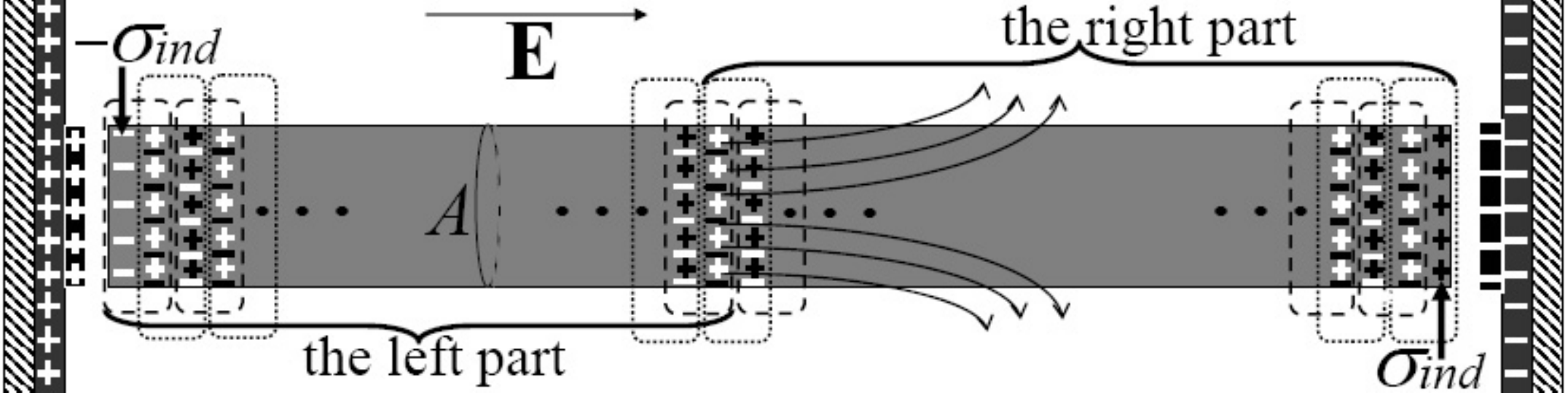}
\caption{A dielectric cylinder in a uniform EF parallel to its
axis is a stack of dipolar layers. Schematic illustration.
\label{fig3}}
\end{center}
\end{figure}

The same is relevant to the interaction between two parts of DB.
(The two parts of DB are to attract each other if the DB is in
equilibrium as a suspension bridge.) A DL cylinder in a uniform
EF~${\bf E}$ (see Fig.~\ref{fig3}) is a stack of same and equally
oriented one-dipolar-molecule-thick double electrostatic layers.
Successive layers penetrate each other: the area of their
overlapping is neutral, since the positive charge of one layer and
the negative charge of the other are intermixed there. They are
schematically distinguished in Fig.~\ref{fig3} by different
rectangles (short and long) and by different colors of charges
(white and black). The positive charge of the last layer at one
base of the cylinder and the negative charge of the last layer at
the other base are not neutralized. Surface densities of these
charges are right equal to the charges induced on the cylinder
bases:~$\sigma_{ind}$ and~$-\sigma_{ind}$. This means that each of
the layers is a~$\pm\sigma_{ind}$ double layer. The left and the
right parts of a long cylinder, each consisting of an integer
number of layers (see Fig.~\ref{fig3}), interact as follows.
(Dividing the cylinder by a plane into two not overlapping parts
would have no sense because dipole molecules of one layer would be
cut into pieces belonging to different parts). The last right
layer (short and white) of the left part and the last left layer
(long and black) of the right part overlap. The left cylindric
part has charge density~$\sigma_{ind}$ induced on its right base,
while the right part has the charge density~$-\sigma_{ind}$
induced on its left base. The EF produced in the right part by the
left part is the EF of the charge~$\sigma_{ind}$ of the left part
right base because its left base is far off. Since the right part
is long it is attracted by the coulomb force~$2\pi\sigma_{ind}^2A$
to the left part like the whole cylinder is attracted to the left
plane in Fig.~\ref{fig2}. At the same time, the overlapping layers
belonging to the two different parts repel each other with the
same force: each of them consists of the~$\sigma_{ind}$
and~$-\sigma_{ind}$ charges, there are six different couples of
these charges, in two of the couples charges repel each other with
the force~$2\pi\sigma_{ind}^2A$, in another couple charges attract
each other with the same force, in one more couple there is no
parallel to the cylinder axis interaction between the charges
because they overlap, the last two couples do not count as they
are the two layers, i.e. rigidly bound parts of molecules. By the
way, the same forces expulse from the cylinder its last layers at
the bases. This is the origin of the difference between~$P$
and~$\tilde P$.

So, the total tension of DB produced by EF is zero if DB is long
enough, and we still have no explanation of the stability. But let
us just estimate the tension of DB produced by {\it surface
tension} (ST). Probably, it has not been done before because there
was a strong belief in the electrostatic origin of DB tension. ST
can hold DB if:
\begin{eqnarray}
\rho g AL \cong 2l \gamma \Theta\label{eq10}
\end{eqnarray}
where~$\gamma$ is the liquid ST,~$\Theta$ is the small angle
between an end of the DB and the horizontal (see Fig.~\ref{fig1}),
and~$l$ is the DB cross-section perimeter. We have supposed that
the WB cross-section is roughly an ellipse~\cite{Marin_2010} with
the height about $1.5$~times larger than the width, and analyzed
using Eq.~(\ref{eq10}) the photos of WBs presented in
Refs.~\cite{Fuchs_neutron,Dynamics,Saija_2010_model,Marin_2010}.
The obtained by us values of the WBs tensions caused by ST are
10-40\% lower than the corresponding ones necessary for holding
the WBs. The discrepancy may be caused by the low accuracy of our
"experimental" investigation or of the approximation in
Eq.~(\ref{eq10}) (one should take into account that the
cross-section of WB is in fact not constant, especially at the
ends). May be also, the reason is that the field between the
beakers is in reality not uniform and therefore it slightly pulls
WB up. Anyway, it is clear that ST is the main force holding DB.
We have also analyzed the photos of glycerine DBs in the setup
with the configuration producing a uniform EF~\cite{Marin_2010}.
If the cross-section of the glycerine DBs is a circle (the side
views only are presented in the Ref.~\cite{Marin_2010}), according
to the Eq.~(\ref{eq10}), the values of the tension are 10\% lower
and 40\% higher than the ones necessary for holding the DBs for
respectively Figs.~7 left and middle in Ref.~\cite{Marin_2010}.

Why is a DB not possible then without electric field? Because ST
plays actually an ambivalent role. On the one hand it does not
allow gravity to tear DB. But on the other hand, as has been
mentioned in Ref.~\cite{Marin_2010}, ST "wants" to break DB into
separate round drops, because then the surface energy would
decrease, i.e.~DB is in a labile equilibrium without the outer
longitudinal electric field. The latter provides the stable
equilibrium: it does not allow the distortion of the DB shape to
start, because the energy of electric field is the lowest if the
shape is nonperturbed. The phenomenon has been extensively studied
long
ago~\cite{Saville_1997,Nayyar_1960,Gonzalez_1989,Gonzalez_1993}.
In Ref.~\cite{Nayyar_1960} energy change caused by small
sinusoidal distortions of an infinite cylindrical jet of DL (an
infinite DB in zero gravity, in other words) have been analyzed.
It has been proved that the longitudinal EF~$E_{cr}$ necessary for
providing the stable equilibrium is~$\sim \sqrt{\gamma}$ and it is
the lower the larger is~$A$ or~$\varepsilon$. In
Ref.~\cite{Gonzalez_1989} the equilibrium shape of a bridge of one
DL surrounded by another DL of the same density has been studied.
Existence of an equilibrium shape very close to the cylindrical
one was used as the instability criterion, and same results have
been obtained: $E_{cr}$ is proportional to the square root of the
ST between the DLs and it is the lower the lower is $L/\sqrt{A}$
or the larger is the ratio of the DLs dielectric constants. Now we
can explain why a long DB is hard to make: it must be thin to stay
the gravity (see Eq.~(\ref{eq10})) but a thinner DB needs a much
stronger field to keep the shape. The model of
Ref.~\cite{Gonzalez_1989} has been generalized in
Ref.~\cite{Gonzalez_1993} for the case when the DB is vertical and
there is a small difference in the two liquids densities. It has
been shown that even the small axial gravity is an important
factor destabilizing the equilibrium between the effects of the
field and ST. This explains the lower stability of vertical WBs as
compared to the horizontal ones~\cite{Ponterio_2010_raman}.

Our speculations describe the basic role of ST and electric field
in providing DB stability. They do not explain why the horizontal
WB cross-section increases with the increase of the voltage
between the beakers~\cite{Ponterio_2010_raman} and why the
horizontal glycerine DB changes its shape~\cite{Marin_2010} if the
external \emph{uniform} EF is altered. In the both cases the
reason may be that EF, even a uniform one, affects the shape of DB
if the shape is asymmetric (it is, actually), and if there are
some free charges in the DL. In the first case nonuniformity of
the EF between the beakers also may play a role.

Latterly, let us propose two small hints for experiment. 1) It has
been reported that WB is possible in an oscillating electric
field~\cite{Ponterio_2010_raman}. At the same time it is known
that the water must be deionized, evidently because free charges
relocate, thus screening the field. But if the field oscillates
frequently enough the ions do not have time to
relocate~\cite{Gonzalez_1989}. May be when using a high frequency
oscillating voltage one does not need to deionize the water. It
would be also possible then to measure the tension of DB and not
the Coulomb attraction of electrodes. 2) It is interesting to make
a DB of a liquid having dielectric permittivity higher than water
has. May be one can obtain then a longer DB. Dielectric constant
of \emph{N}-Methylformamide (NMF) is around
200~\cite{NMF_diel_permittivity,Swiergiel_2009}. The challenge is
to make sure that NMF is really free of ion-producing
contaminations: of water first of all. Otherwise the conductivity
is too high~\cite{Swiergiel_2009}.

We thank~E. Fuchs and K.~Gatterer for introducing the WB
phenomenon and providing their pioneering works on the subject;
V.I. Denisov, O.E. Philippova, A.V. Gorshkov, A.M. Lotonov, I.A.
Malyshkina, S.V.~Venev, and A.F.~Aerov for discussions.



\end{document}